\documentclass[trackchanges]{aastex701}
\usepackage{multirow}
\usepackage{graphicx}
\usepackage{booktabs}
\usepackage{makecell}
\usepackage{bm}
\usepackage{subfigure}
\usepackage{amsmath,amssymb}
\usepackage[T1]{fontenc}
\graphicspath{{./}{figures/}}
\usepackage{longtable}
\usepackage{cases}
\usepackage{epstopdf}
\usepackage[]{amsmath}
\usepackage{threeparttablex}
\usepackage{mathtools}

\begin{document}

\title{SN2023syz and SN2025cbj: Two Type IIn Supernovae Associated with IceCube High-energy Neutrinos}

\author[orcid=0000-0002-9067-3828]{Ming-Xuan Lu}
\affiliation{Guangxi Key Laboratory for Relativistic Astrophysics,
	School of Physical Science and Technology, Guangxi University, Nanning 530004,
	China}
\affiliation{GXU-NAOC Center for Astrophysics and Space Sciences, Nanning 530004, People's Republic of China}
\email{lumx@st.gxu.edu.cn}  

\author[orcid=0000-0002-6316-1616]{Yun-Feng Liang}
\affiliation{Guangxi Key Laboratory for Relativistic Astrophysics,
	School of Physical Science and Technology, Guangxi University, Nanning 530004,
	China}
\affiliation{GXU-NAOC Center for Astrophysics and Space Sciences, Nanning 530004, People's Republic of China}
\email[show]{liangyf@gxu.edu.cn}

\author[orcid=0000-0001-8411-8011]{Xiang-Gao Wang}
\affiliation{Guangxi Key Laboratory for Relativistic Astrophysics,
	School of Physical Science and Technology, Guangxi University, Nanning 530004,
	China}
\affiliation{GXU-NAOC Center for Astrophysics and Space Sciences, Nanning 530004, People's Republic of China}

\email[show]{wangxg@gxu.edu.cn}

\author{Hao-Qiang Zhang}
\affiliation{Guangxi Key Laboratory for Relativistic Astrophysics,
	School of Physical Science and Technology, Guangxi University, Nanning 530004,
	China}
\affiliation{GXU-NAOC Center for Astrophysics and Space Sciences, Nanning 530004, People's Republic of China}
\email{2007301181@st.gxu.edu.cn}

\begin{abstract}
Type IIn supernovae (SNe IIn) are a subclass of core-collapse SNe in which strong interactions occur between the ejecta and dense circumstellar material, creating ideal conditions for the production of high-energy neutrinos. This makes them promising candidate sources of neutrinos.
In this work, we conduct an association study between 163 SNe IIn observed by the Zwicky Transient Facility and 138 neutrino alert events detected by the IceCube neutrino observatory. After excluding alerts with poor localization, we find two SNe that are spatiotemporally coincident with neutrino events. IC231027A and IC250421A coincide with the positions of SN2023syz and SN2025cbj, respectively, within their localization uncertainties, and the neutrino arrival times are delayed by 38 days and 61 days relative to the discovery times of the corresponding SNe. 
Using Monte Carlo simulations, we estimate that the probability of such two coincidences occurring by chance in our sample is $p \sim 0.67\%$, suggesting that they may originate from genuine physical associations, though the result is not yet statistically significant. Our model calculations, however, indicate that the likelihood of a neutrino originating from IC231027A is low, implying that the association between IC231027A and SN2023syz is likely coincidental. Nevertheless, under optimistic parameters, the probability of detecting a neutrino from the whole SNe IIn sample could reach $\gtrsim6\%$, indicating that detecting neutrino emission from the SNe population may be possible. Our study provides a systematic analysis, combining statistical analysis and model calculations, to assess whether interacting supernovae can serve as potential sources of neutrino emission.
\end{abstract}


\section{Introduction} 
IceCube's observations confirmed the existence of high-energy (TeV–PeV) neutrinos originating from the cosmic space \citep{IceCube:2013low,IceCube:2014stg,IceCube:2016umi}, but the sources of these diffuse neutrinos remain not clear. In recent years, some candidate neutrino sources have been detected by IceCube, such as the blazar TXS 0506+056 and the Seyfert galaxy NGC 1068 \citep{IceCube:2018dnn,IceCube:2018cha,Aartsen:2019fau,icecube2022evidence}. Neutrino emissions from the galactic plane have also been detected with a significance of 4.6$\sigma$ \citep{icecube2023observation}. Some other observational evidence also supports that Seyfert galaxies are a promising class of neutrino-emitting objects \citep{Neronov:2023aks,IceCube:2024dou,IceCube:2024ayt,Sommani:2024sbp}. However, the origins of most of the diffuse neutrino flux can still not be fully explained. Some promising candidates, such as blazars and radio AGNs, have been found to account for at most 30\% of the total flux \citep{IceCube:2016qvd,Hooper:2018wyk,Smith:2020oac,Yuan:2019ucv,Zhou:2021rhl,2022PhRvD.106h3024L}. Therefore, further searches for the sources of high-energy neutrino emission are necessary. Numerous studies have already been conducted targeting different types of astrophysical objects, including pulsar wind nebulae (PWN) \citep{2020ApJ...898..117A}, X-ray binaries \citep{IceCube:2022jpz,Fang:2024wmf}, active galactic nuclei (AGN) \citep{Halzen:1997hw,IceCube:2016qvd,Hooper:2018wyk,Smith:2020oac,Yuan:2019ucv,Zhou:2021rhl}, gamma-ray bursts (GRB) \citep{Waxman:1997ti,Abbasi:2009ig,2012ApJ...752...29H,Aartsen:2014aqy}, and tidal disruption events (TDE) \citep{2021NatAs...5..510S,2022PhRvL.128v1101R,2023ApJ...948...42W,2024ApJ...969..136Y,2024MNRAS.529.2559V,2024arXiv241106440L,Lu:2025vmk}.


In addition to serving as sources of MeV neutrino emission (as has been shown by Supernova 1987A \citep{Kamiokande-II:1987idp}), supernovae have also been extensively studied as potential sources of TeV–PeV neutrinos \citep{Zirakashvili:2015mua,Murase:2017pfe,He:2018lwb,Li:2018tfh,Chang:2022hqj,IceCube:2023esf,Cosentino:2025sdd}.
Among all types of supernovae, Type IIn SNe are a subclass of core-collapse supernovae (CCSNe) whose spectra exhibit strong, narrow hydrogen emission lines (“n” standing for “narrow”). These narrow lines, often accompanied by long-lasting and luminous light curves, originate from the collision of high-velocity supernova ejecta with very dense circumstellar material (CSM) shed by the progenitor star shortly before explosion \citep{2017hsn..book..403S}.
Such intense interaction between the supernova ejecta and the dense CSM creates ideal conditions for high-energy neutrino production. 
As the fast-moving supernova ejecta plows into the dense CSM, powerful shocks can form, accelerating protons to ultra-high energies \citep{Murase:2010cu}.
These high-energy protons then collide with the dense CSM through proton–proton ($pp$) interactions, producing muons which in turn decay into neutrinos \citep{Murase:2010cu,Murase:2017pfe,Kheirandish:2022eox,Pitik:2023vcg}.
Because having both an powerful particle accelerator and a dense target, Type IIn supernovae possess the two key ingredients needed for neutrino production, making them stand out among other types of supernovae as promising sources detectable by the IceCube neutrino observatory.

Models show that nearby SNe IIn (tens of Mpc or closer) with efficient CR acceleration could produce a neutrino signal detectable by IceCube \citep{Petropoulou:2017ymv}.
As a population they might contribute a non-negligible fraction of IceCube’s diffuse flux under optimistic assumptions \citep{Petropoulou:2017ymv}.
Single-source and stacking searches conducted with IceCube have yielded meaningful constraints, but so far there is no definitive and widely accepted detection of a Type IIn supernova as a high-energy neutrino source \citep{IceCube:2023esf}.
However, follow-up observations of neutrino alerts do find a few interacting SNe candidates (interacting SNe in general including Ibn/IIn) \citep{IceCube:2015jsn,Stein:2025uxi}.
Recently, \citet{Stein:2025uxi} found an interacting supernova, SN 2023uqf, in the optical follow-up observations of IceCube alert events with the Zwicky Transient Facility (ZTF), which coincided in time with the high-energy neutrino IC231004A,
providing an observational evidence that interacting supernovae can serve as hadronic cosmic-ray accelerators.


In view of the above arguments, a systematic search for associations between SNe IIn and neutrinos is necessary. 
This will help determine whether SNe IIn are effective high-energy neutrino emitters and quantify their contribution to the IceCube diffuse neutrino flux.
In this work, we conduct a systematic search for spatial-temporal coincidences between SNe IIn and IceCube alert events.
We use the SNe IIn classified by the ZTF - Bright Transient Survey (BTS) \citep{Perley:2020ajb} as our supernova sample. This is a public catalog with a strict pipeline to conduct a classification for transient sources, and it conducts a magnitude-limited ($m < 19\,{\rm mag}$ in either the $g$ or $r$ filter) survey for extragalactic transients in the ZTF public stream \citep{Perley:2020ajb}.
We find two spatial-temporal coincidence events, namely IC231027A-SN2023syz ($z$ = 0.037) and IC250421A-SN2025cbj ($z$ = 0.0675).
We calculate the model-predicted number of neutrino events for these two sources to check whether they are compatible with observations.

\section{Association analysis} \label{sec:style}
\subsection{The Neutrino Sample}
High-energy astrophysical neutrinos beyond the atmospheric background have been observed by IceCube in both cascade and track data. 
To help determine the origin of these neutrinos through follow-up observations, since 2016, alerts of individual high-energy neutrino events have been released in real time to platforms for the multi-messenger observation community, such as the General Coordinates Network (GCN)\footnote{\url{https://gcn.gsfc.nasa.gov/amon_icecube_gold_bronze_events.html}}. These alert events mainly focus on track-like neutrino candidates, as they have higher angular resolution than cascade events.
In 2019, the alert system was updated. Each alert event is now assigned a “signalness” value (representing the probability that the alert is of astrophysical origin). Based on this value, alerts are categorized as “gold” or “bronze,” corresponding to astrophysical origin probabilities greater than 50\% and 30\%, respectively (assuming a spectral index of –2.19 for astrophysical neutrinos). 
For each alert event, information on energy, direction, the 90\% position uncertainty ($r_{90}$, statistical error only), and arrival time is reported. For each alert event, updated reports will be issued after further processing provides more accurate information. In our analysis, we use the latest report information for each alert event.
We note that one alert event (IC241016\footnote{\url{https://gcn.nasa.gov/circulars/37794}}) was explicitly determined not to be of astrophysical origin, and is therefore excluded from our sample.

We note that \citet{IceCube:2023agq} has provided alert events from before 2019. However, considering the start time of the ZTF operation, we focus primarily on neutrinos released after 2019, since this period better matches the survey time of the ZTF-BTS catalog. This allows alert events to occur after the SNe listed in the ZTF-BTS catalog. Therefore, for consistency, in this analysis we use only alerts listed in the GCN after 2019. We include all bronze and gold alert events in our sample, which have at least a 30\% probability of astrophysical origin.
In addition, in the association analysis, a good spatial localization of neutrino events is essential, as large localization errors can lead to a high rate of chance coincidences. Many previous studies (e.g., \cite{Plavin:2020emb,Hovatta:2020lor}) excluded alert events when the area of their 90\% confidence level (C.L.) error region (denoted as $\Omega_{90}$) exceeds a certain threshold. Accordingly, in this analysis, we retain only alert events with $\Omega_{90} \leq 30\,\text{deg}^2$ to remove those alerts with large error regions. Our final alert sample contains a total of 138 alert events from June 19, 2019 to June 30, 2025, as shown in Fig.~\ref{fig: sample}.

\subsection{The SNe IIn Sample from ZTF-BTS}
\begin{figure*}
\center
\includegraphics[width=0.8\textwidth]{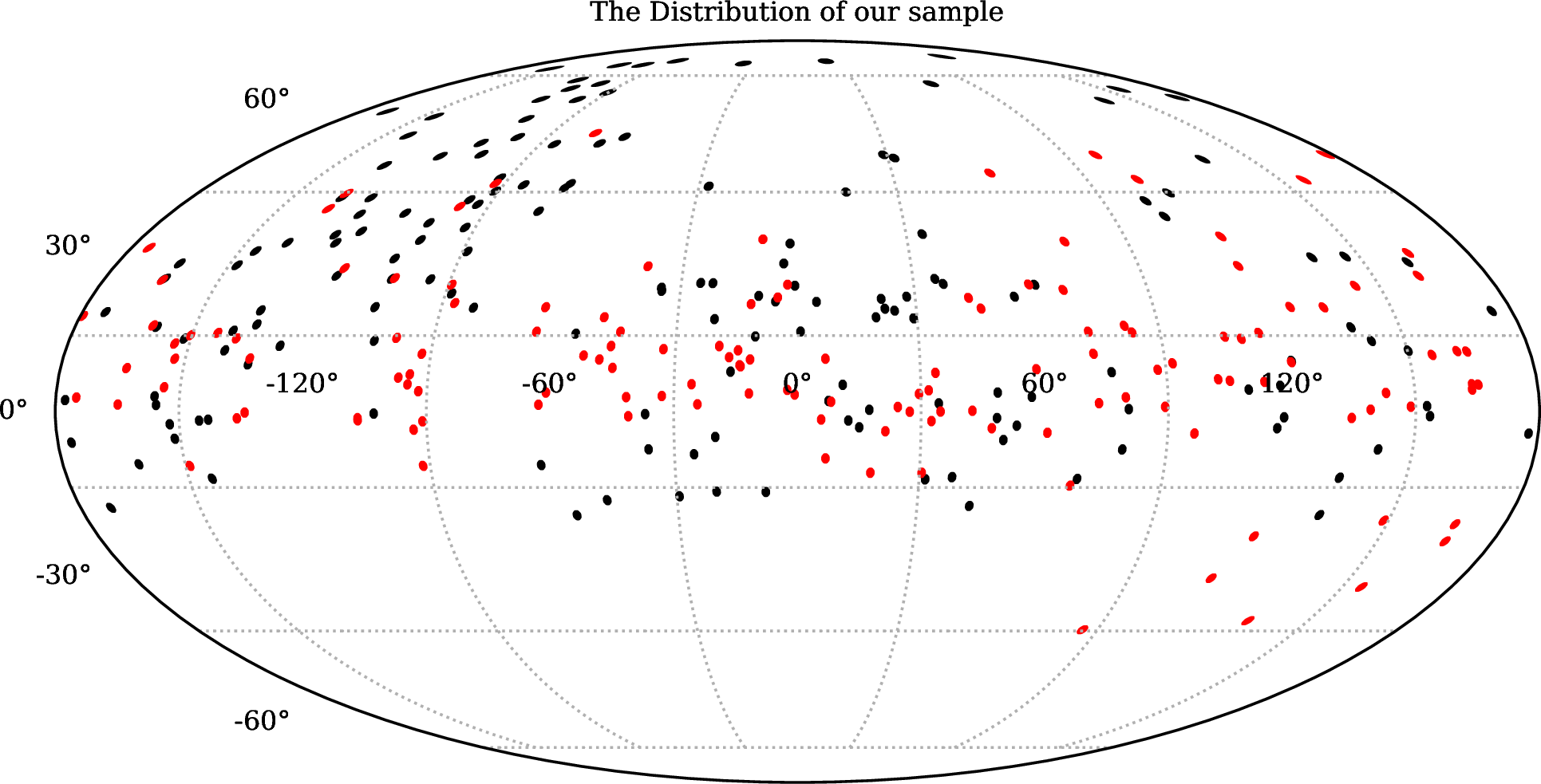}
\caption{Sky distributions in equatorial coordinates of the SNe IIn and alert events adopted in this work. The black points represent the positions of 163 SNe IIn, and the red points represent the locations of 138 alert events.}
\label{fig: sample}
\end{figure*}

ZTF is a optical time-domain survey that uses a wide-field (47 deg$^2$) camera on the Samuel Oschin Telescope at Palomar Observatory to systematically scan the night sky \citep{2019PASP..131a8002B}. Its primary purpose is to discover transient sources.
Since 2018, ZTF has conducted a survey of the visible northern sky ($\sim 3\pi$) with a three-day cadence, and announce newly discovered transient candidates via public alerts \citep{Masci:2018neq,Graham:2019qsw,2019PASP..131a8002B,Fremling:2019dvl}. In December 2020, ZTF entered the second phase (Phase-II) of its public survey operations, with the cadence increased to a two-day cadence.
In addition to its photometric survey, ZTF-BTS carries out an extensive spectroscopic campaign aimed at spectroscopically classifying all extragalactic transients with peak magnitudes brighter than 18.5 mag in the $g$ or $r$ bands (and, when spectroscopic resources allow, transients as faint as magnitude 19.0 mag). All classification results are publicly released \citep{Fremling:2019dvl}. The survey has cataloged more than 10,000 objects and is updated nearly daily\footnote{\url{https://sites.astro.caltech.edu/ztf/bts/bts.php}}.
The ZTF-BTS catalog adopts a set of strict quality criteria for transient classification, such as magnitude limit, number of observations, and Galactic extinction threshold (see \citet{Perley:2020ajb} for details). 
Consequently, this source catalog provides a good sample of SNe IIn/candidates with reliable classification for performing the association analysis. 
Based on the classifications reported in the ZTF-BTS, we select a total of 163 SNe IIn as of June 30, 2025. We consider only sources with high classification reliability exhibiting ``SN-like behavior'' (slow rise and/or fade time, or coincident with a galaxy). The sky distribution of these SNe is shown in Fig.~\ref{fig: sample}. 

\subsection{Crossmatch and Chance Probability}
We perform a systematic search for SNe IIn coincident with neutrino events based on the following criteria: (1) the position of the SN IIn lies within the $r_{90}$ error radius of the neutrino event; (2) the arrival time of the neutrino alert event is within 100 days of the discovery date of the SN IIn. 
We find that SN2025cbj and SN2023syz are spatially and temporally associated with the neutrino events IC250421A and IC231027A, respectively (see Table~\ref{table:properties} for their information). 
IC250421A and IC231027A arrived 61 and 38 days after the discovery date of SN 2025cbj and SN 2023syz (2025 February 19 and 2023 September 19, MJD $=$ 60725 and 60206), respectively. 
The angular distances between the SNe and the central positions of their corresponding alert events are 1.79$^\circ$ for SN 2025cbj and 2.10$^\circ$ for SN 2023syz. 
We note that the spatial coincidence between IC250421A and SN2025cbj has already been reported in \citet{2025GCN.40208....1G}. For IC231027A, no reports of an SN association is found; but two Fermi-LAT sources were identified within a small angular distance (< 0.3°) of the best-fit position of IC231027A \citep{2023GCN.34932....1B}. We also perform a Fermi-LAT data analysis of these two sources, as described in Appendix~\ref{Fermi-LAT analysis}. In addition, \citet{2025GCN.40242....1P} found no Fermi-LAT detection at both the position of SN2025cbj and the best-fit location of IC250421A and provided upper limits on the gamma-ray emission above 100 MeV from these two positions.

\begin{table}[ht]
\caption{\label{table:properties}The properties of SN2023syz and SN2025cbj.}
\begin{center}
\hspace{-2.5cm}
\begin{tabular}{ccccccccc}
\toprule
 & $t_{\rm rise, obs}^a$ & $M_{\rm abs}^b$ & Redshift & RA & DEC & Association & $E_\nu$ &Time delay\\
 & [days] & & & [deg] & [deg] & & [TeV] & [days] \\
\toprule
SN2023syz & 10& $-17.58$ &0.037 & 268.85 &45.22 & IC231027A & 191.5 &38\\
SN2025cbj& 50 & $-19.15$ &0.0675 & 239.92 &27.11 & IC250421A & 151.4 & 61 \\
\toprule
\end{tabular}
\begin{tablenotes}
\item $^a$ {The time interval between the first detection time and the peak time of the SNe.}
\item $^b$  The absolute magnitude for these two sources, which can serve as indicators of the source’s emission power.
\end{tablenotes}
\end{center}
\end{table}
We then estimate the chance probability of obtaining two coincidence events using Monte Carlo (MC) simulations. Since IceCube is located at the South Pole and its detection sensitivity depends only on the zenith angle \citep{IceCube:2016tpw,Aartsen:2019fau}, we generate mock neutrino samples by randomizing the right ascension (RA) while keeping the declination (DEC) fixed at their observed values. 
The arrival times of the mock neutrinos are uniformly drawn between the earliest and latest arrival times in the dataset. In addition, we generate simulated SN IIn samples by randomizing their positions within a radius of $10^\circ$ around their original locations\footnote{ The $10^\circ$ choice of the SNe IIn simulation follows previous works \citep[e.g.,][]{Buson:2022fyf,Plavin:2022oyy,Buson:2023irp,wm72-25tq}, which ensure sufficient randomization while preserving the actual spatial distribution. We have also tested additional radii ranging from 4$^{\circ}$ to 20${^\circ}$, finding that this parameter only has a small impact on the results.}, while requiring that each simulated source remain within the sky coverage of the ZTF-BTS field. We also apply a Kolmogorov–Smirnov (KS) test to the distributions of $\rm RA$ and $\rm DEC$ for the simulated sources to ensure that the mock samples preserve the distribution characteristics of the real sources. The mock source catalogs generated in this way has almost identical spatial distribution as the real catalog (see Apendix~\ref{app:kstest}).

By performing $10^4$ Monte Carlo simulations, we estimate the chance probability (i.e., the $p$-value for rejecting the null hypothesis) of finding at least two matches with a time delay of less than 61 days. The probability was computed using $p = {(M+1)}/{(N+1)}$ where $M$ is the number of simulations in which the statistic (i.e., the number of matches) is greater than or equal to the observed value, and $N=10^4$ is the total number of MC simulations \citep{Davison_Hinkley_1997}. We obtain a chance probability of $p \sim 0.67\%$.
This result reflects that the probability of occurring two (or more) coincidence in our sample by chance is low. 
It is likely that the SNe and neutrino associations we have found are physically real, but the current results have not yet reached a level of statistically significant.

\section{Model-predicted neutrino counts from SN 2025cbj and SN 2023syz} \label{sec:floats}

The statistical analysis above provides supporting evidence for SNe IIn being sources of high-energy neutrinos. 
Below, we will investigate whether such a result aligns with model expectations.

We consider a spherically symmetric ejecta propagating into the CSM, with ejecta mass $M_{\rm ej}$ and kinetic energy $E_k$. The CSM is assumed to be spherical, steady, and wind-like, with an outer radius $R_{\rm csm}$ and a total mass $M_{\rm csm}$. The number density of the CSM can be expressed as $n_{\rm CSM}(R) = {\dot{M}}/{4 \pi v_{w} m R^{2}}$, where $\dot{M}$ is the mass-loss rate of the progenitor star, $v_{w}$ is the stellar wind velocity, and $m = \mu m_{\rm p}$. Here $m_{\rm p}$ is the proton mass, $\mu = 1.3$ is the mean molecular weight of a neutral gas of solar abundance, and $R$ is the distance from the stellar core.
The CSM is shed by the progenitor star before the SN explosion. The stellar wind velocity is taken to be $v_{w} = 100\,\rm km\,s^{-1}$ \citep{Moriya:2014cua}, and the average mass-loss rate is \citep{Suzuki:2020yhz}
\begin{equation}
\dot{M} = 0.3\ {M}_{\odot}\ {\rm yr}^{-1} 
\left(\frac{M_{\rm CSM}}{10\ M_{\odot}}\right)
\left(\frac{R_{\rm CSM}}{10^{16}\ \rm cm}\right)^{-1}
\left(\frac{v_{w}}{100\ \rm km\ s^{-1}}\right).
\end{equation}

The interaction between the ejecta and the CSM creates shocks that accelerate particles. In our calculation, we only consider the forward shock and neglect the contribution from the reverse shock following previous works \citep[e.g.,][]{Petropoulou:2016zar, Pitik:2021dyf}. The temporal evolution of the radius of the shocked shell could be described as \citep{Suzuki:2020qui, Chevalier:2016hzo, Moriya:2013hka, Pitik:2021dyf}
\begin{subnumcases}{R_{\rm {sh}}(t) =}
\Bigg[\frac{2}{s(s-4)(s-3)}\frac{10(s-5)E_{\rm{k}}}{3(s-3)M_{\rm {ej}}}\frac{v_w}{\dot{M}}\Bigg]^{1/(s-2)} \times t^{(s-3)/(s-2)},  & \mbox{for} $R \leq R_{\rm dec}$ \\
R_{\rm {dec}}\left(\frac{t}{t_{\rm {dec}}}\right)^{2/3}, & \mbox{for} $R > R_{\rm dec}$
\label{eq1:R_sh}
\end{subnumcases}
where $R_{\rm dec} = {M_{\rm ej} v_w}/{\dot{M}}$ is the deceleration radius, and $s = 10$ is the outer slope of the ejecta density profile. Consequently, the shock velocity is $v_{\rm {sh}}(t) = dR_{\rm sh}/dt \propto t^{(-1)/(s-2)}$ for $R \leq R_{\rm dec}$ and $v_{\rm {sh}}(t) \propto t^{-1/3}$ for $R > R_{\rm dec}$.

During the expansion of the shock, protons are accelerated, and high-energy neutrino emission can be produced through interactions between shock-accelerated protons and cold protons in the dense CSM ($pp$ interaction) \citep{2012IAUS..279..274K, IceCube:2021xar}. At radii smaller than the shock breakout radius $R_{\rm bo}$, where the optical depth satisfies $\tau > c/v_{\rm sh}$, the high-density environment suppresses acceleration. Therefore, efficient particle acceleration only occurs after this radius \citep{1976ApJS...32..233W, Levinson:2007rj, Murase:2010cu}. The shock breakout radius can be determined by solving
\begin{equation}
\tau_{T}(R_{\rm{bo}})=\int_{R_{\rm{bo}}}^{R_{\rm{CSM}}} \rho_{\rm{CSM}}(R)\,\kappa_{\rm{es}}\, dR = \frac{c}{v_{\rm{sh}}} ,
\end{equation}
where $\kappa_{\rm{es}}\sim 0.34\,\rm{cm}^{2} \rm{g}^{-1}$ \citep{Pan:2013nfa} is the electron scattering opacity at solar abundance, $c$ is the speed of light, and $\rho_{\rm CSM}$ is the CSM density at radius $R$.

The evolution of the proton distribution between $R_{\rm bo}$ and $R_{\rm CSM}$ follows \citep{1997ApJ...490..619S, 2012ApJ...751...65F, Petropoulou:2016zar}:
\begin{equation}
\frac{\partial N_{\rm{p}}(\gamma_{\rm{p}},R)}{\partial R} - \dfrac{\partial }{\partial \gamma_{\rm{p}}}\bigg[\frac{\gamma_{\rm{p}}}{R} N_{\rm{p}}(\gamma_{\rm{p}}, R)\bigg] + \frac{N_{\rm{p}}(\gamma_{\rm{p}}, R)}{v_{\rm{sh}}(R)\, t_{pp}(R)} = Q_{\rm{p}}(\gamma_{\rm{p}},R),
\label{eq:diffeq}
\end{equation}
where $N_{\rm{p}} (\gamma_{\rm{p}}, R)$ denotes the number density of protons at a given radius $R$ with energies between $\gamma_{\rm{p}}$ and $\gamma_{\rm{p}} + \rm{d}\gamma_{\rm{p}}$. The second term on the left-hand side represents energy losses due to the adiabatic expansion of the SN shell, while $pp$ collisions are treated as an effective escape process, with a characteristic timescale $t_{\rm pp} = (k_{\rm{pp}} \sigma_{\rm {pp}} n_{\rm {sh}} c)^{-1}$ \citep{1997ApJ...490..619S}.

For a wind density profile of CSM, the proton injection rate in Eq.~(\ref{eq:diffeq}) is \citep{Pitik:2021dyf}
\begin{equation}
Q_{\rm p}(\gamma_{\rm p}, R) = \frac{9\pi\epsilon_{\rm p}R^2_{\rm bo}n_{\rm bo}}{8{\ln}\left({\gamma_{p,\rm{max}}}/{\gamma_{p,\rm{min}}}\right)} \Bigg[\frac{v_{\rm sh}(R_{\rm bo})}{c} \Bigg]^2\Bigg[\frac{R}{R_{\rm bo}} \Bigg]^{2\alpha}\gamma_{\rm p}^{-k}, \quad \text{for}\ \ \gamma_{p,\rm {min}} < \gamma_{p}< \gamma_{p,\rm {max}},
\label{eq1:R_sh} 
\end{equation}
where $\alpha$ is the radial dependence index of the shock velocity, $\alpha = -1/7$ ($-1/2$) for $R < R_{\rm dec}$ ($R > R_{\rm dec}$). $\gamma_{p,\rm{min}}$, $\epsilon_p$, $k$ are the minimum Lorentz factor of protons, the proportion of kinetic energy converted into proton acceleration, and the proton spectral index, respectively. We follow the previous works of, e.g., \citet{Pitik:2021dyf,Murase:2010cu}, and set them as ($\gamma_{p,\rm{min}}$, $\epsilon_p$, $k$) = ($1, 0.1, 2$). The maximum Lorentz factor of protons $\gamma_{p,\rm max}$ can be calculated by requiring that the acceleration timescale is shorter than the cooling timescale. For proton cooling, we consider the cooling due to inelastic collisions and adiabatic expansion of the shocked shell. More details on the calculation of $\gamma_{p,\rm max}$ can be seen in \citet{Pitik:2021dyf}.

The accelerated protons produce neutrinos through $pp$ interactions. After solving $N_p(\gamma_p,R)$ from Eq.~(\ref{eq:diffeq}), combining it with the cross section of $pp$ interaction $\sigma_{\rm pp}$, the neutrino production rate ($Q_{\nu_{i} + \bar{\nu}_{i}}$ with $i = \mu, e$) can be given by .
\begin{equation}
Q_{\nu_{i} + \bar{\nu}_{i}}(E_{\nu}, R) = \frac{4 n_{\rm{CSM}}(R) m_{\rm{p}} c^{3}}{v_{\rm{sh}}} \int_{0}^{1} {\rm d}(\ln x) {\sigma_{\rm{pp}}(E_{\nu}/x)} N_{\rm{p}} \left(\frac{E_{\nu}}{x m_{\rm{p}} c^{2}}, R\right)F'_{\nu_i}(E_{\nu}, x)
\label{eq: neutrino rate}
\end{equation}
where $x = E_{\nu}/E_{\rm p}$ and $F'_{\nu_i}(E_{\nu}, x)=F^{(1)}_{\nu_{\mu}}(E_{\nu}, x) + F^{(2)}_{\nu_{\mu}}(E_{\nu}, x)$ for $i=\mu$ and $F'_{\nu_i}(E_{\nu}, x)=F_{\nu_{e}}(E_{\nu}, x)$ for $i=e$. The $F^{(1)}_{\nu_{\mu}}$, $F^{(2)}_{\nu_{\mu}}$, and $F_{\nu_{e}}$ adopt the form in \citet{Kelner:2006tc}, which are valid for $E_p > 0.1$ TeV. 

The observed neutrino and antineutrino flux of each flavor ($F_{\nu_{\alpha}+\bar{\nu}_{\alpha}}$ with $\alpha = e,\mu, \tau$) can be calculated as 
\begin{equation}
\label{eq:F}
F_{\nu_{\alpha}+\bar{\nu}_{\alpha}}(E_{\nu}, t) = \frac{(1 + z)^2}{4 \pi d^{2}_{L}(z)} v_{\rm{sh}}(t)\underset{\beta}{\sum} P_{\nu_{\beta} \rightarrow \nu_{\alpha}}Q_{\nu_{\beta} +\bar{\nu}_{\beta}}\left(E_{\nu_{\alpha}}(1 + z),R(t)\right),
\end{equation}
where $P_{\nu_{\beta} \rightarrow \nu_{\alpha}}$ is the flavor change due to neutrino oscillations during the propagation of neutrinos \citep{Anchordoqui:2013dnh,Pitik:2021dyf}.

Finally, the expected event rate of muon neutrinos detected by IceCube is 
\begin{equation}
\label{eq: neutrino event number}
\dot{N}_{\nu_{\mu}+\bar{\nu}_{\mu}}(t)= \int_{E_{\nu,\rm{min}}}^{E_{\nu,\rm{max}}} dE_{\nu} A_{\rm{eff}}(E_{\nu}, \delta)  F_{\nu_{\mu}+\bar{\nu}_{\mu}}(E_{\nu}, t)
\end{equation}
where $A_{\rm eff} (E_\nu, \delta)$ is the detector effective area for which we use the effective area given in \citet{IceCube:2023agq}.
For a given DEC bin, the dependence of $A_{\rm eff}$ on neutrino energy is shown in Figure 3 of \citet{IceCube:2023agq}. We determine $E_{\nu,\rm min}$ based on the lower boundary of the effective area curve. For SN2023syz (DEC = 45.22$^\circ$) and SN2025cbj (DEC = 27.11$^\circ$), the corresponding values of $E_{\nu,\rm min}$ are 1.0 TeV and 1.4 TeV, respectively.

$E_{\nu,\rm max}$ should have the same evolutionary profile as that of $E_{\rm p, max}$, and it has lower energy than $E_{\rm {p, max}}$.
The value of $E_{\rm p, max}$ varies with the shock radius/time. Before and after $R_{\rm dec}$, it follows $E_{\rm p, max} \propto R^{4/7} \propto t^{1/2}$ and $E_{\rm p, max} \propto R^{-1/2} \propto t^{-1/3}$, respectively \citep{Pitik:2021dyf}. This indicates that $E_{\rm p, max}$ and $E_{\nu,\rm max}$ peak at $R_{\rm dec}$.
For a given set of SN model parameters (i.e. $M_{\rm ej}$, $M_{\rm CSM}$, $R_{\rm CSM}$, and $E_k$), the evolution of $E_{\nu,\rm max}$ is determined by the relative ordering of $R_{\rm dec}$, $R_{\rm bo}$, and $R_{\rm CSM}$\footnote{The parameter space for the SN model should be constrained to cases satisfying $R_{\rm bo} <R_{\rm CSM}$, since efficient particle acceleration occurs only after $R_{\rm bo}$.}.
If $R_{\rm bo} <R_{\rm dec}< R_{\rm CSM}$, $E_{\nu,\rm max}$ initially increases until the shock reaches $R_{\rm dec}$, after which it decreases. In the cases $R_{\rm dec} <R_{\rm bo}< R_{\rm CSM}$ and $R_{\rm bo} < R_{\rm CSM}<R_{\rm dec}$, $E_{\nu,\rm max}$ would continuously decrease and increase, respectively, throughout the shock evolution ranging from $R_{\rm bo}$ to $R_{\rm CSM}$.

To cover the broad parameter space of SNe IIn \citep{Pitik:2021dyf,Pitik:2023vcg,Ransome:2024cza}, we explore a wide range of values for ($M_{\rm ej}$, $M_{\rm CSM}$, $R_{\rm CSM}$). Based on ZTF-BTS observations, the adopted $E_k$ ranges are $(1-8)\times$ 10$^{50}$ erg for SN2023syz and $(0.3-3) \times$ 10$^{52}$ erg for SN2025cbj. It should be noted that, for SN2025cbj, the adopted kinetic energy values may be relatively optimistic, although it seems they can still align with the limited existing observations (see Appendix \ref{Ek choice} for details).

In Figure~\ref{fig1: events rate}, we present the neutrino event rate of SN2023syz and SN2025cbj for chosen sets of model parameters.
By integrating the temporal interval from the breakout time $t_{\rm {bo}}$ to the arrival dates of alert events $t_f$, i.e. $t_{\rm bo} = 60206$ (MJD), $t_f =60244$ (MJD) for SN2023syz and $t_{\rm bo} =60725$ (MJD),  $t_f = 60786$ (MJD) for SN2025cbj, the total number of neutrinos can be calculated by
$N_{\nu_{\mu}+\bar{\nu}_{\mu}}= \int_{t_{\rm{bo}}}^{t_f}dt\ \dot{N}_{\nu_{\mu}+\bar{\nu}_{\mu}}(t)$.



\begin{figure*}[htbp]
\center
\subfigure[1-100 TeV for SN2023syz]{
\includegraphics[width=0.45\textwidth]{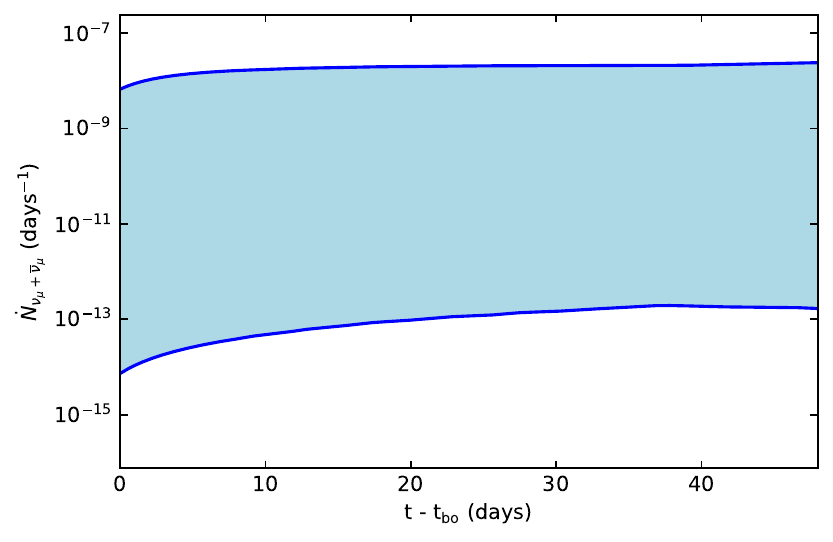}}
\subfigure[0.1-10 PeV for SN2023syz]{
\includegraphics[width=0.45\textwidth]{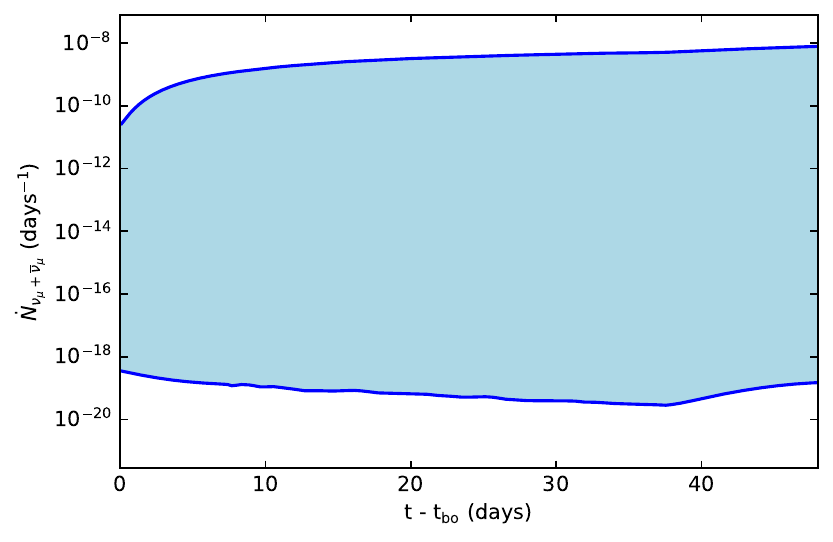}}
\subfigure[1-100 TeV for SN2025cbj]{
\includegraphics[width=0.45\textwidth]{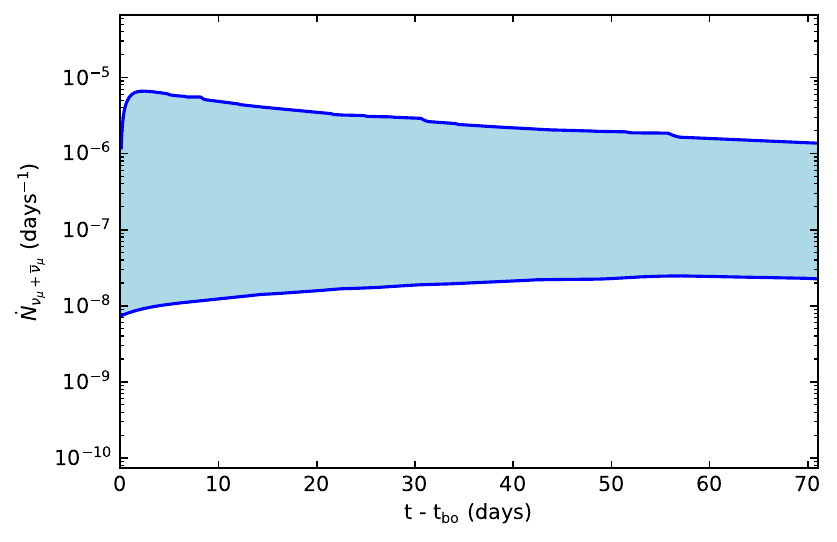}}
\subfigure[0.1-10 PeV for SN2025cbj]{
\includegraphics[width=0.45\textwidth]{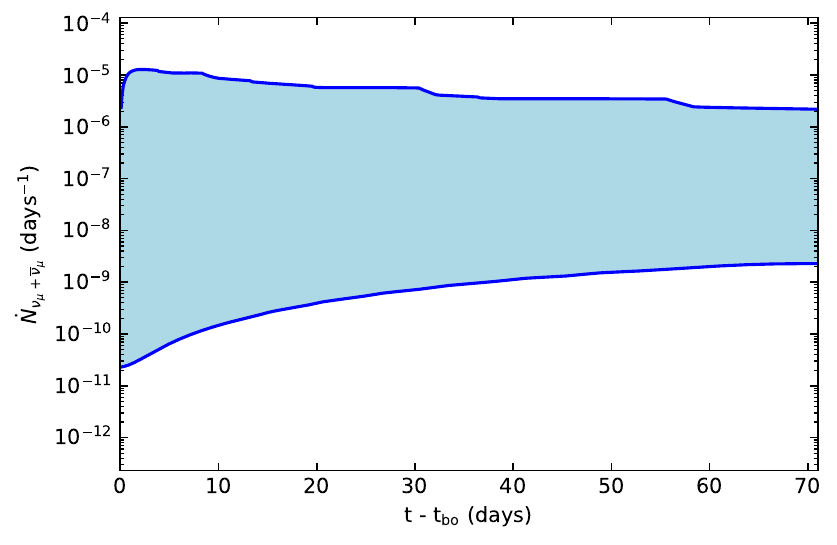}}
\caption{Neutrino event rates integrated over different energy ranges as a function of the time after the shock breakout for SN2023syz and SN2025cbj, respectively. The results assume the following parameters sets: $(E_k, R_{\rm CSM}) = (8\times10^{50} \rm erg, 4\times10^{16} cm)$  for SN2023syz and $(E_k, R_{\rm CSM}) = (3\times10^{52} \rm erg, 4\times10^{16} cm)$ for SN2025cbj, respectively. The blue band denotes the variation arising from varying the $M_{\rm ej}$ and $M_{\rm CSM}$  parameters in the range of ($M_{\rm ej}$, $M_{\rm CSM}$) $\in$ ($[1 - 20]\,M_{\odot}$, $[1 - 30]\,M_{\odot}$) for SN2023syz and 
($M_{\rm ej}$, $M_{\rm CSM}$) $\in$ ($[1 - 70]\,M_{\odot}$, $[1 - 70]\,M_{\odot}$) for SN2025cbj. When deriving the blue band, we have excluded the parameters not satisfying $t_{\rm rise}\leq t_{\rm rise,obs} \leq1.5  t_{\rm rise}$ and $t_{\rm CSM}-t_{\rm bo}$ greater than the observed time delay, please see the main text and Figure~\ref{fig: total events} for details.}
\label{fig1: events rate}
\end{figure*}

\begin{figure*}
\center
\subfigure[1-100 TeV for SN2023syz]{
\includegraphics[width=0.45\textwidth]{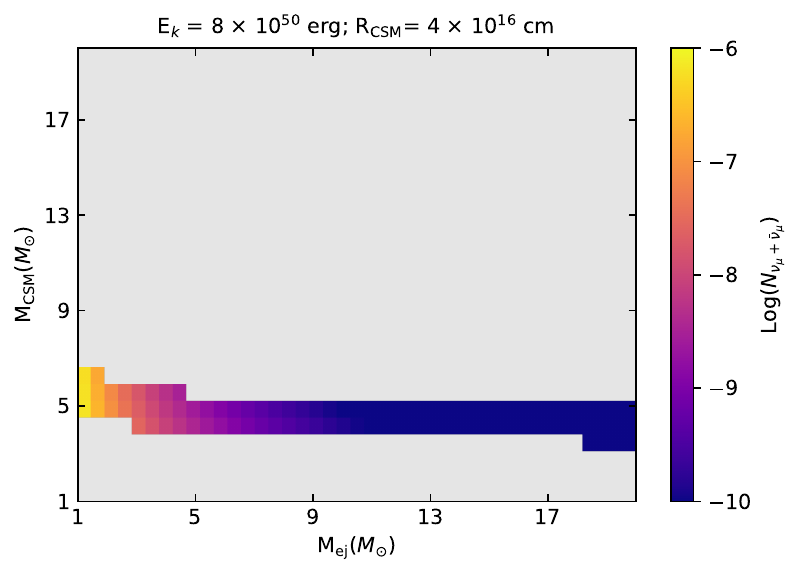}}
\subfigure[0.1-10 PeV for SN2023syz]{
\includegraphics[width=0.45\textwidth]{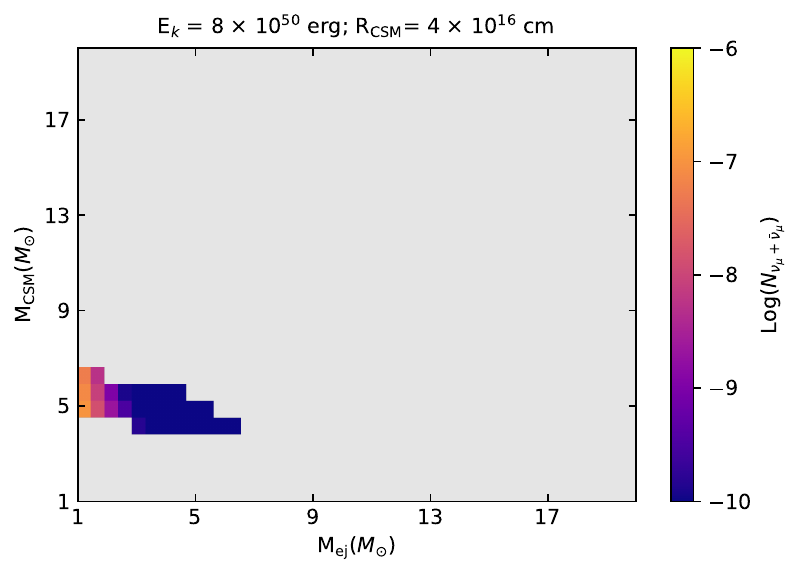}}

\subfigure[1-100 TeV for SN2025cbj]{
\includegraphics[width=0.45\textwidth]{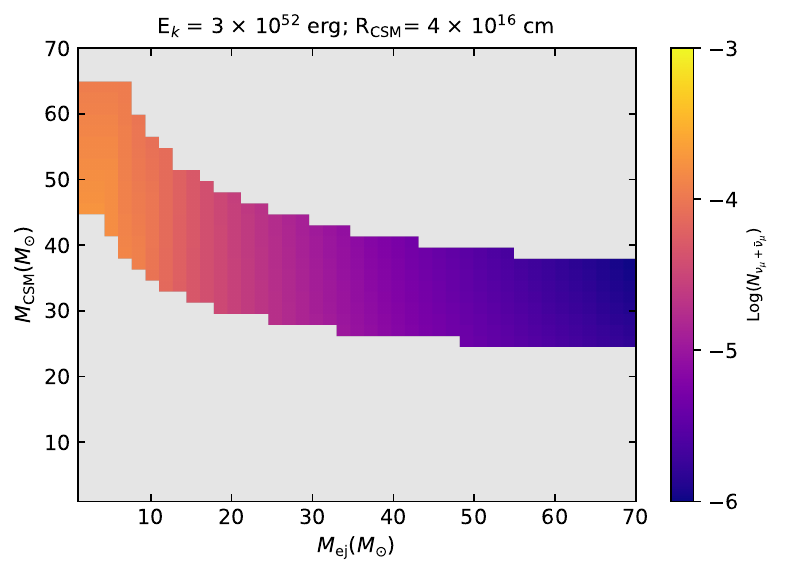}}
\subfigure[0.1-10 PeV for SN2025cbj]{
\includegraphics[width=0.45\textwidth]{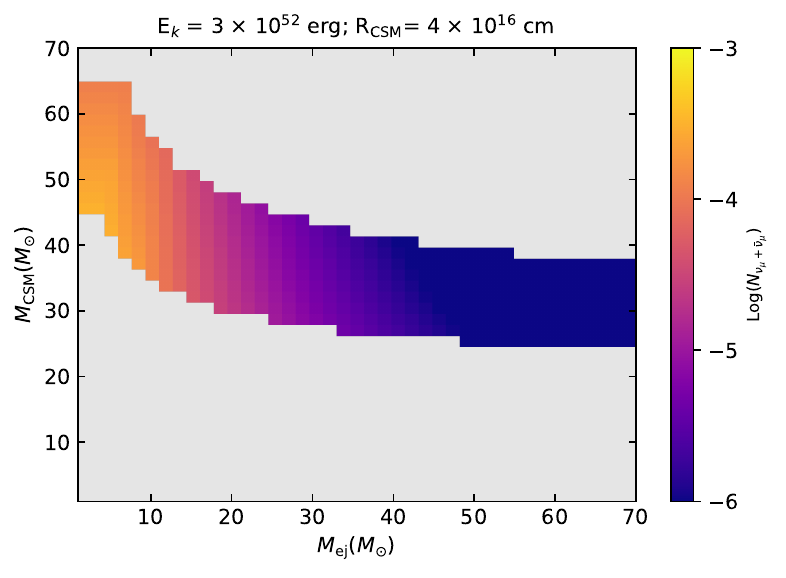}}
 
\caption{Expected total number of neutrinos, integrated over different energy ranges, from SN 2023syz (upper panels) and SN 2025cbj (lower panels).
We scan the parameters of ($M_{\rm ej}$, $M_{\rm CSM}$) = ($[1 - 20]\,M_{\odot}$, $[1 - 20]\,M_{\odot}$); ($M_{\rm ej}$, $M_{\rm CSM}$) = ($[1 - 70]\,M_{\odot}$, $[1 - 70]\,M_{\odot}$) to show how the expected number relies on these parameters for SN2023syz and SN2025cbj, respectively. 
The gray shaded region in the figure represents the excluded parameters for which the expected $t_{\rm rise}$ and $t_{\rm {CSM}} - t_{\rm{bo}}$ do not match the observations. 
}
\label{fig: total events}
\end{figure*}

Figure~\ref{fig: total events} presents the model-predicted total number of muon (anti)neutrino events detected by IceCube from SN2023syz and SN2025cbj. In the figure, we have excluded those parameters (shaded region) in the parameter space according to the $t_{\rm {rise}}$ (the time interval between the first detection time\footnote{Note that using the first detection time is a conservative choice, as emission is usually present before it.} and the peak time of the SNe) and $t_{\rm {CSM}} - t_{\rm{bo}}$ (the time delay between $t_{\rm{bo}}$ and $t_{\rm {CSM}}$), where $t_{\rm {CSM}}$ is the time shock reaching the radius $R_{\rm CSM}$. 
According to the light curves of SN2023syz\footnote{\url{https://alerce.online/object/ZTF23abfglcy}} and SN2025cbj\footnote{\url{https://alerce.online/object/ZTF25aagbrpb}}, we adopt the $t_{\rm{rise, obs}}$ of 10 days and 50 days for SN2023syz and SN2025cbj, respectively. The $t_{\rm{rise}}$ can be linked to the other quantities through the following relation \citep{2012ApJ...757..178G,Pitik:2023vcg},

\begin{equation}
\label{Eq: t_diff}
  t_{\rm{rise}} \sim \int_{R{_{\rm bo}}}^{R_{\rm ph}} \frac{2(R-R_{\rm bo})k_{\rm T}\rho_{\rm CSM}(R) {\rm d}R}{c}
\end{equation}
where the $R_{\rm {ph}}$ is the photosphere radius at the optical depth $\tau_{\rm{T}}(R_{\rm ph})$ = 1. Then we require the parameter space to satisfy $t_{\rm{rise}}\leq t_{\rm{rise,obs}} \leq 1.5\,t_{\rm{rise}}$ \citep{Pitik:2023vcg}, namely we allow an error up to 50\% on the $t_{\rm{rise}}$ estimated from the model.
Furthermore, the $t_{\rm CSM}-t_{\rm bo}$ is required to be larger than the time delay between the arrival dates of alert events and SNe IIn (i.e. $t_{\rm CSM}-t_{\rm bo}\geq$ 38 days for SN2023syz and 61 days for SN2025cbj, respectively). This criterion ensures that the forward shock is still interacting with the CSM at the moment the alert event occurred.

In Figure~\ref{fig: total events}, we can see that the expected number of detected neutrinos depends on the values of $M_{\mathrm{ej}}$ and $M_{\mathrm{CSM}}$.
For SN 2023syz, its narrow parameter range is mainly due to the short rise time $t_{\mathrm{rise}}$ of this SN.
Within the considered parameter space, the most optimistic choice of the parameters (i.e., the yellow part in the parameter space) yield $N_{\nu_{\mu}+\bar{\nu}_{\mu}} \sim 1.8\times10^{-4}$ and $3.1\times10^{-4}$ for SN2025cbj
by integrating 1-100 TeV and 0.1-10 PeV energy ranges, respectively. For SN2023syz, the corresponding values are $N_{\nu_{\mu}+\bar{\nu}_{\mu}} \sim $ 1.1$\times 10^{-7}$ and 6.7$\times 10^{-7}$, respectively.
These results indicate that the detection probability of neutrino events from a single SN IIn is very low, especially when the neutrino kinetic energy is small.
The association between IC231027A and SN2023syz is likely coincidental.

Next, we estimate the expected detection probability of neutrinos from the entire SN IIn population.
We assume that the expected neutrino number for each source is related to the IceCube effective area $A_{\rm eff}$ (which is related to the DEC and energy of alerts) and the supernova's peak flux ($F_{\rm peak}$) or luminosity distance ($1/d_L^2$):
\begin{equation}
\label{Eq: tot_neutrino}
N_{\rm expect} \propto \int_{E_{\nu,1}}^{E_{\nu,2}} {A_{\rm eff}(E_\nu, {\rm DEC})}\times X \, dE_\nu.
\end{equation}
with $X$ being $F_{\rm peak}$ or $1/d_L^2$.
The reason for us to adopt peak flux and distance as the weighting factors in the calculation is that the former can serve as a proxy for the SN's kinetic energy, while the latter assumes that the distribution of kinetic energies is the same for SNe at different distances.
Using the  mean value of $N_{\nu_{\mu}+\bar{\nu}_{\mu}}$ obtained from SN2023syz and SN2025cbj (approximately $2.5 \times 10^{-4}$ per source, since the expected neutrino counts from SN2023syz is negligible), we estimate that in the optimistic scenario, the total expected number of neutrinos from 163 SNe IIn is $\sim$0.06 (0.11) when $X$ is $F_{\rm peak}$ ($1/d_L^2$), with the contributions from the energy ranges 1–100 TeV and 0.1–10 PeV approximately 0.03 (0.04) and 0.03 (0.07), respectively. Our model calculations show that, under optimistic parameter choices, there is a certain probability ($\sim5\%$) to detect one neutrino alert event. Therefore, the observed neutrino event in the direction of SN2025cbj originating from this supernova is still possible.



\section{Summary}

This study presents a systematic search for spatial-temporal associations between ZTF Type IIn supernovae and high-energy neutrino alerts detected by the IceCube neutrino observatory.
SNe IIn are core-collapse supernovae characterized by strong interactions between their ejecta and dense CSM. These interactions create ideal conditions for neutrino production through efficient proton acceleration and $pp$ interactions, making SNe IIn promising candidates for neutrino sources detectable by IceCube.
In view of this, we search for spatial and temporal coincidences between 163 SNe IIn observed by ZTF-BTS and 138 IceCube neutrino alerts (2019$-$2025) with well localization ($\Omega_{90} \leq 30\,\rm deg^2$).

Our analysis identify two SNe-neutrino coincidences: SN2023syz ($z = 0.037$) with neutrino event IC231027A (38-day delay after supernova onset) and SN2025cbj ($z = 0.0675$) with IC250421A (61-day delay). The later is also reported in \citet{2025GCN.40208....1G}. Both neutrinos arrived within the localization uncertainty of their respective supernovae. Monte Carlo simulations estimate that the chance probability of observing $\geq2$ such coincidences in our sample is only $\sim$0.67\%, suggesting these associations may be physical real rather than by chance, though not yet statistically conclusive.
To assess the physical plausibility of these associations, we perform model calculations of neutrino emission from these two sources.
Using a neutrino production model involving ejecta evolution, shock acceleration, and $pp$ interactions, we compute the expected muon neutrino event rates at IceCube, accounting for CSM properties (mass $M_{\rm CSM}$, radius $R_{\rm CSM}$), ejecta kinetic energy ($E_{k}$) and mass ($M_{\rm ej}$), and observational constraints from the rise time $t_{\rm rise}$ of the optical light curves. 
For SN2025cbj, the predictions based on the most optimistic parameter sets explored in this work (the yellow regions in Figure~\ref{fig: total events}) indicate that the number of detectable muon neutrino events, $N_{\nu_{\mu}+\bar{\nu}_{\mu}}$, is $\sim1.8 \times 10^{-4}$ in the 1–100 TeV band and $\sim3.1 \times 10^{-4}$ in the 0.1–10 PeV band. The expected number of neutrino events from SN2023syz is estimated to be at the $10^{-7}$ level and is negligible, making its association with IC231027A likely a chance coincidence rather than physically real. When considering the most optimistic parameter set, our calculation shows that the total expected number of neutrinos detectable from the sample of 163 Type IIn supernovae is 0.06 $\sim$ 0.11 (depending on how different SNe are weighted in the calculation), and the detection of one neutrino coincidence is still possible. It should also be noted that when more common or typical (i.e., not that optimistic) SN parameters are adopted, the expected number of neutrinos may significantly decrease, making the detection of neutrinos from SNe much less likely.

This work provides important insights into assessing whether interacting supernovae, particularly SNe IIn, could be potential sources of high-energy neutrinos. 
Our analysis shows a relatively low probability of chance coincidence. Although model calculations indicate a low expected number of neutrino events, detection remains possible under suitable parameters, and the physical association between SN2025cbj and IC250421A cannot yet be ruled out.
Type II superluminous supernovae (SLSNe-II) exhibit higher emission power and ejecta kinetic energy. If the detected association were physically real, one would expect more neutrino events to be observed from SLSNe. However, the current absence of any observed associations between SLSNe and neutrino alerts poses a challenge to such a model scenario. We leave the investigation of SLSNe to future work. Future observations with enhanced neutrino sensitivity and more powerful transient surveys (e.g., Vera Rubin Observatory \citep{2019ApJ...873..111I,2020ApJ...890...73B}) will help confirm or rule out the associations between SNe IIn and neutrinos definitively and quantify the role of SNe IIn in cosmic neutrino production.

\begin{acknowledgments}
This work is supported by the National Key Research and Development Program of China (Grant No. 2022YFF0503304), the National Natural Science Foundation of China (12373042, U1938201, 12494573), the Programme of Bagui Scholars Programme (WXG) and Innovation Project of Guangxi Graduate Education (YCBZ2024060).
\end{acknowledgments}

\bibliography{sample701}{}
\bibliographystyle{aasjournalv7}
\clearpage
\begin{appendix}

\section{KS-test in the Monte Carlo simulation}
\label{app:kstest}
\begin{figure*}[h]
\centering
\includegraphics[width=0.4\textwidth]{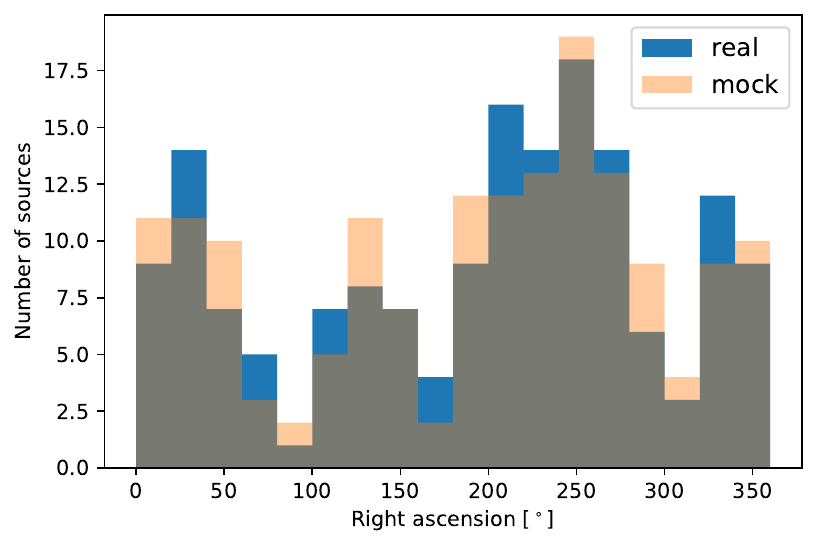}
\includegraphics[width=0.4\textwidth]{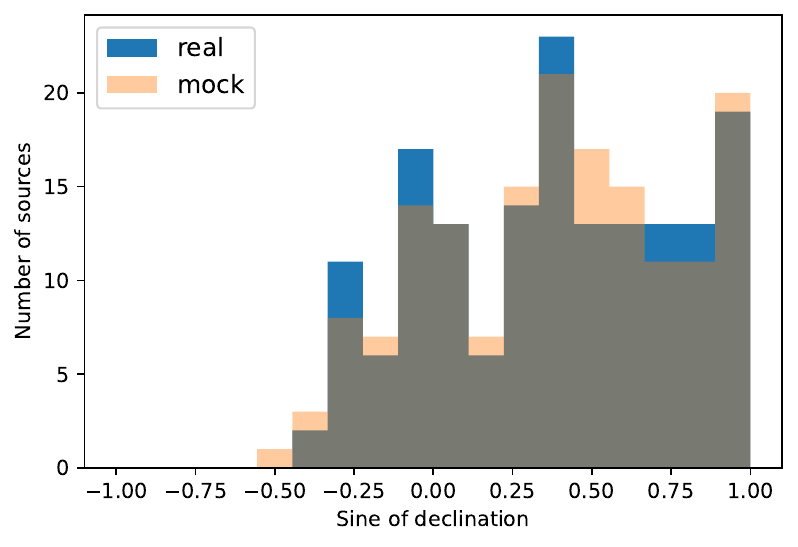}
\caption{A comparison of the RA and DEC distributions between the real SN IIn catalog and one realization of the mock catalogs.}
\label{fig:distr}
\end{figure*}

In the Monte Carlo simulation, we apply a KS test to the distributions of $\rm RA$ and $\rm DEC$ for the simulated sources to ensure that the mock samples preserve the distribution characteristics of the real sources. {The comparison of spatial distribution between mock sources and real ones after applying the KS test is presented in Figure~\ref{fig:distr} for a demonstration. The mock source catalog generated in this way has almost identical spatial distribution as the real catalog.}

\section{\textit{Fermi}-LAT data analysis around IC231027A}
\label{Fermi-LAT analysis}
\begin{figure}
\center
\includegraphics[width=0.6\textwidth]{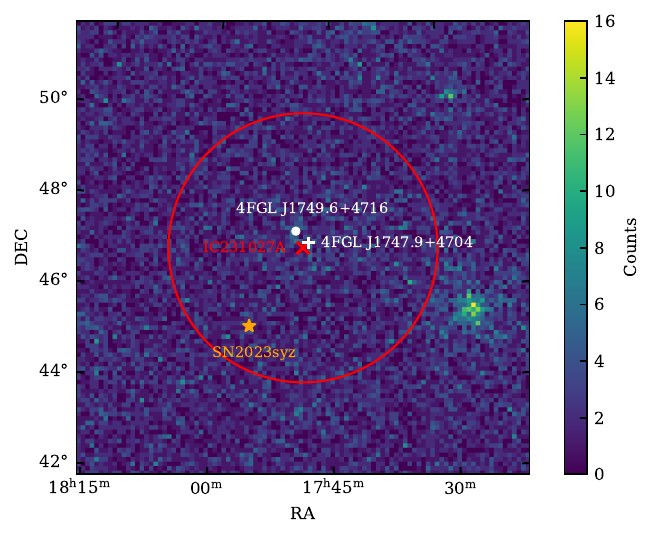}
\caption{This figure is the \textit{Fermi}-LAT counts map in the 100 MeV - 1 TeV energy range, showing the integrated search in the period of $\pm$ 120 days of the arrival date of IC231027A. The position of IC231027A is marked by the red 'X' with the 90\% positional uncertainty denoting by the red circle (only statistical error). The orange star is the position of SN2023syz. The white circle and plus sign are the positions of two 4FGL sources within the error circle of IC231027A (i.e. 4FGL J1749.6+4716 and 4FGL J1747.9+4704), away from the best fit position of IC231027A 0.16 deg and 0.29 deg, respectively.}
\label{fig: cmap}
\end{figure}

The gamma-ray sources of the fourth \textit{Fermi} catalog (4FGL-DR4, for Data Release 4)\footnote{\url{https://fermi.gsfc.nasa.gov/ssc/data/access/lat/14yr_catalog/}}\citep{Fermi-LAT:2019yla,Fermi-LAT:2022byn} has two sources located nearby the IC231027A ($< 0.3^\circ$), as reported by \citep{2023GCN.34932....1B}. We take the \textit{Fermi}-LAT data to analyze these two sources. Centered on the best fit position of IC231027A, we use Pass 8 SOURCE events ({\textit evclass=128, evtype=3}) with energies ranging from 100 MeV to 1 TeV, and select a $15^{\circ} \times 15^{\circ}$ Region of Interest (ROI). The P8R3\_SOURCE\_V3 of LAT instrument response functions are adopted. The Galactic diffuse and isotropic component use the file of \textit{gll\_iem\_v07.fits} and \textit{iso\_P8R3\_SOURCE\_V3\_v1.txt}, respectively. We exclude events with zenith angle $z_{\rm max} < 90^{\circ}$ to reduce the contamination produced by the Earth's limb. We first make a TS map integrating $\pm$ 120 days of arrival date of IC231027A for which find gamma-ray counts nearby the IC231027A. The significant gamma-ray counts are not found during the time span, as shown in figure~\ref{fig: cmap}. Then, for these two 4FGL-DR4 sources as reported in \citet{2023GCN.34932....1B}, we perform a likelihood analysis, with spatial bins of 0.1$^\circ$ and 20 logarithmically-space bins for energy. We adopt the spatial model and spectrum as recorded in the model file of \textit{gll\_psc\_v32.xml}. 

\begin{figure*}
\center
\includegraphics[width=0.45\textwidth]{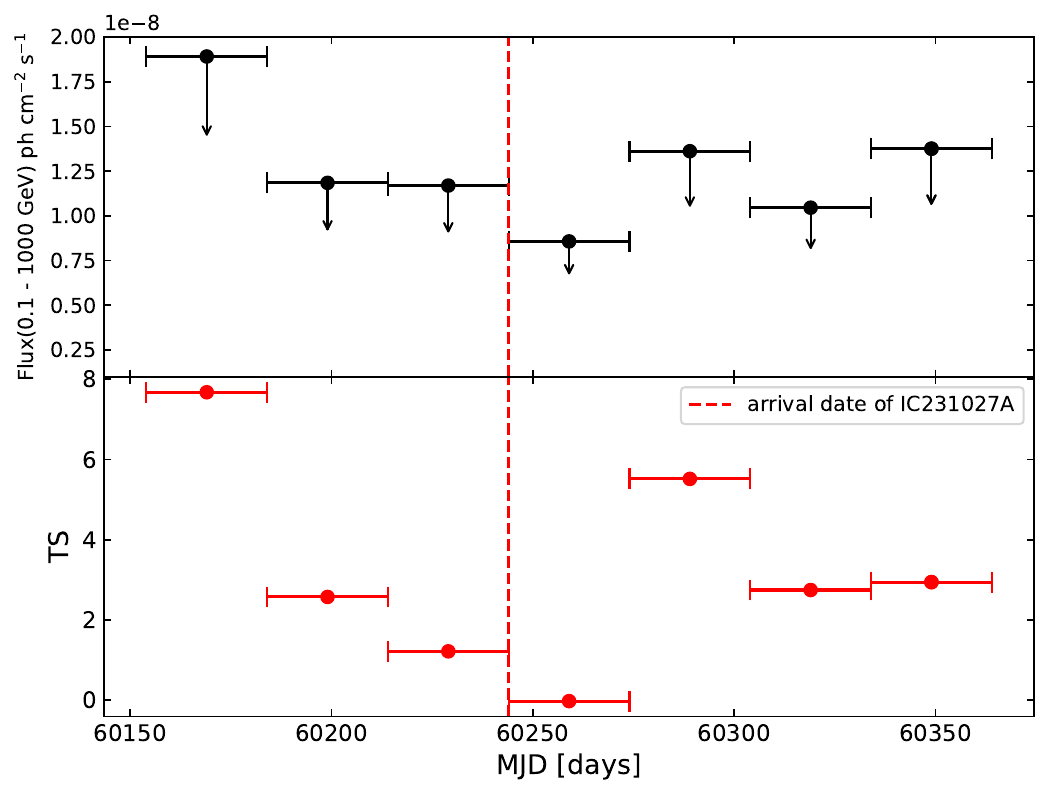}
\includegraphics[width=0.45\textwidth]{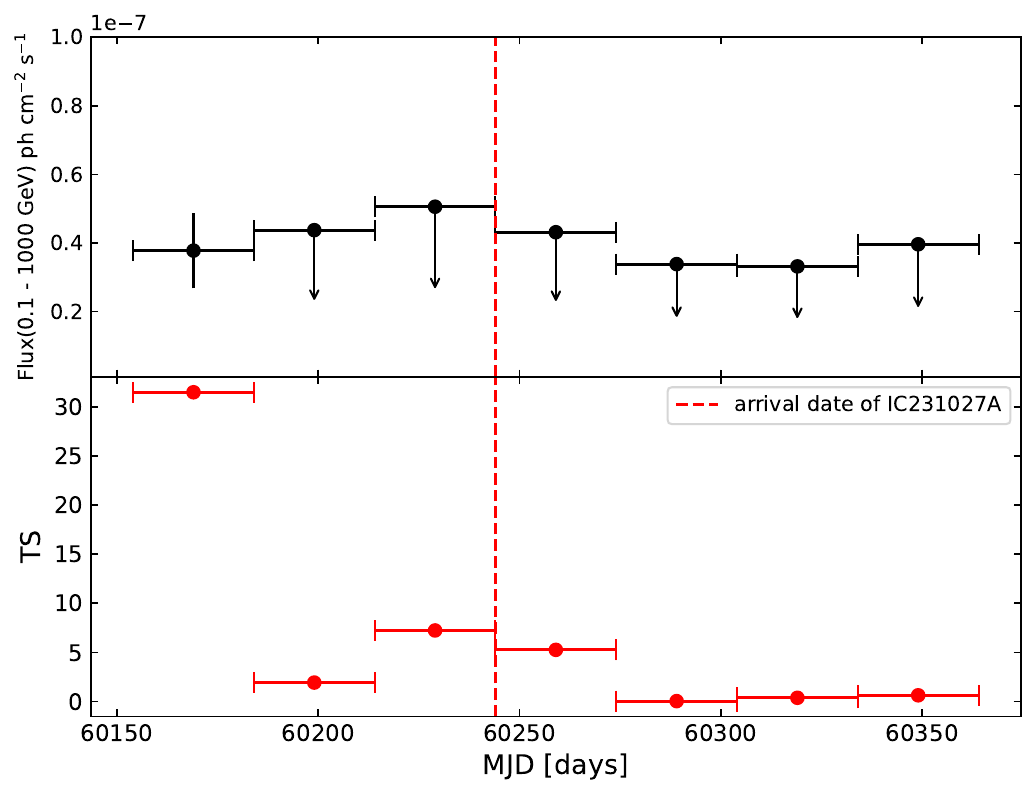}
\caption{\textit{Fermi}-LAT light curves of the two 4FGL-DR4 sources (i.e. 4FGL J1749.6+4716 and 4FGL J1747.9+4704) near the best position of IC231027A ($<0.3^\circ$). We integrate the energies from 100 MeV to 1 TeV. The time spans 90 days before the arrival date of IC231027A and 120 days after it with a 30-day bin. We plot upper limits for points with ${\rm TS}<9$. We don't find any significant gamma-ray signal around the arrival date of IC231027A.}
\label{fig: LC}
\end{figure*}

We present the light curves for these two 4FGL-DR4 sources in figure~\ref{fig: LC}, where upper limits are shwon for time bins with TS<9. The light curves span a period from 90 days before to 120 days after
the arrival date of IC231027A, with a bin size of 30 days. The obvious correlation is not found between \textit{Fermi}-LAT light curves and arrival time of IC231027A. Therefore, we conclude that these two \textit{Fermi}-LAT sources are unlikely to be associated with IC231027A.

\section{The choice of kinetic energy}
\label{Ek choice}

\begin{figure}
\center
\includegraphics[width=0.45\textwidth]{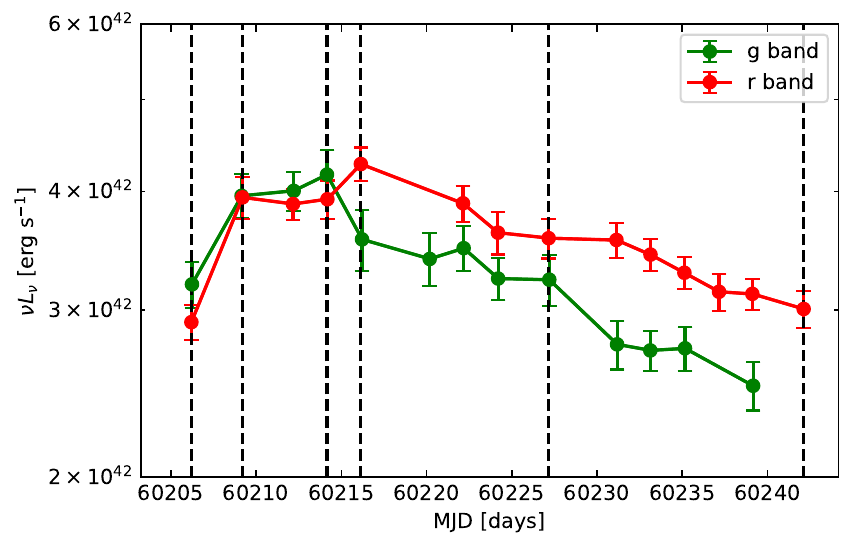}
\includegraphics[width=0.45\textwidth]{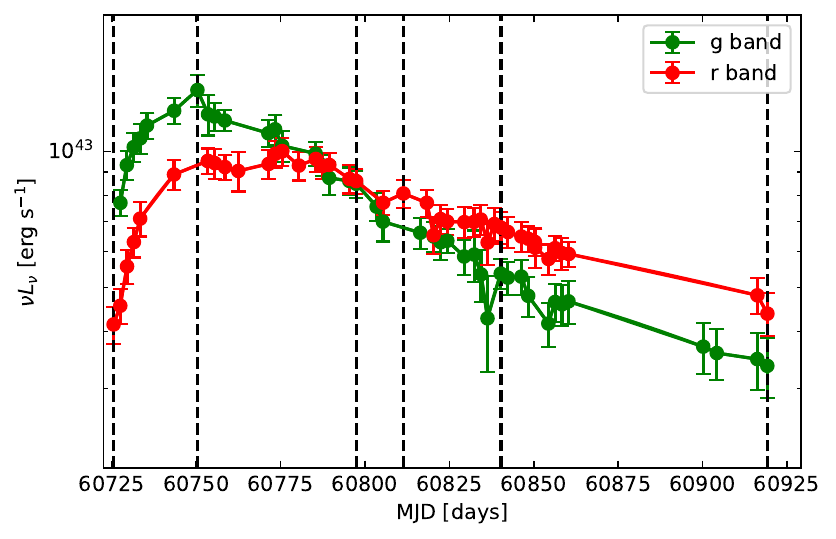}
\caption{Left panel: the ZTF-BTS light curve of SN2023syz. The red and green points are the $r$ band and $g$ band luminosity, respectively. The dashed black lines are the six epochs fitted by the blackbody model (see details in Appendix~\ref{Ek choice}). Right panel: the same as the left panel, but for SN2025cbj.}
\label{fig: obs}
\end{figure}

\begin{figure*}
\center
\includegraphics[width=0.6\textwidth]{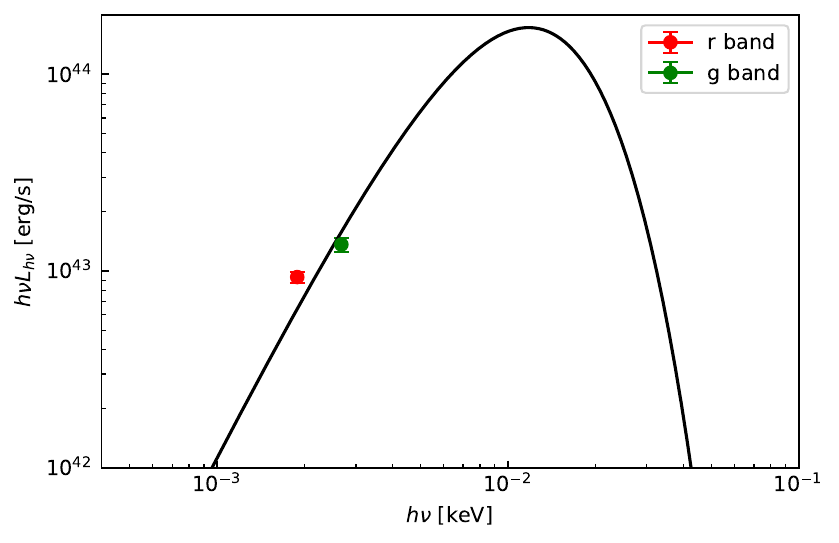}
\caption{Fit of the observational data of SN2025cbj with a blackbody spectral model. The black line corresponds to the model parameters of $(kT, L_{\rm bb})\sim (3\,{\rm eV}, 2.3 \times 10^{44}\,{\rm erg/s})$. The red and green points represent the $r$-band and $g$-band data, respectively, from the second epoch (i.e., the peak epoch) of SN2025cbj.}
\label{fig: bb res}
\end{figure*}

Given the range of radiative efficiency, the $E_k$ could be evaluated an approximate range from the total radiated energy ($E_{\rm rad}$), following the approach of \citet{Pitik:2021dyf, Pitik:2023vcg}. However, the lack of multi-wavelength observations for these two sources restricts the direct evaluation on $E_{\rm rad}$. Therefore, we utilize a blackbody model applied to the ZTF-BTS data to roughly estimate their $E_{\rm rad}$.
The blackbody model\footnote{The blackbody model is described as \url{https://heasarc.gsfc.nasa.gov/docs/software/xspec/manual/node136.html}} is characterized by the temperature ($kT_{\rm bb}$, where $k$ is the boltzmann constant) and the blackbody luminosity ($L_{\rm bb}$). We constrain the values of these two parameters by fitting the data points obtained from ZTF-BTS observations at six epochs.
The continuous light curves of SN2023syz and SN2025cbj are generated using linear interpolation of the data in the $r$ and $g$ band, as show in figure~\ref{fig: obs}. For instance, the peak luminosity of SN2025cbj could be described by the blackbody model with parameters ($kT_{\rm bb}$, $L_{\rm bb}$) = (3 eV , 2.3 $\times$ 10$^{44}$ erg/s), as shown in figure~\ref{fig: bb res}.

We fit these light curves at six epochs as follows. The first epoch is the initial observation. The second epoch is the peak luminosity. Then the third epoch is where the the $r$-band luminosity has the same as the $g$-band one. The fourth and the fifth epochs are selected from periods where the $r$-band luminosity exceeds the $g$-band luminosity. The final epoch corresponds to the last data point. We present these epochs by the dashed black lines in figure~\ref{fig: obs}. However, note that the parameters constraints are limited by the availability of only two bands from ZTF-BTS.


The estimated parameter ranges for the blackbody components are as follows.
For SN2023syz, the $kT$ and $L_{\rm bb}$ are the 0.5-3.1 eV and (0.41-5.27) $\times$ 10$^{43}$ erg/s, respectively.
For SN2025cbj, the $kT$ and $L_{\rm bb}$ are the 0.5-4.3 eV and (0.06-2.52) $\times$ 10$^{44}$ erg/s, respectively.
We can estimate $E_k$ for SN2023syz and SN2025cbj based on assumed mean blackbody temperature of $kT \sim 2$ eV and $kT \sim 3$ eV, respectively. Incorporating the uncertainty in radiative efficiency ($E_{\rm rad}$ = 0.1 $\sim$ 0.8 $E_k$)~\citep{Pitik:2023vcg}, the derived $E_k$ ranges are (1-8) $\times$ 10$^{50}$ erg for SN2023syz and (0.3-3) $\times$ 10$^{52}$ erg for SN2025cbj.

Furthermore, we also search for X-ray counterparts to SN2023syz and SN2025cbj based on the Living \textit{Swift} XRT Point Source Catalog (LSXPS)\footnote{\url{https://www.swift.ac.uk/LSXPS/}} covering the energy range of 0.3-10 keV \citep{Evans:2022bmg}. This search yields no detections within $0.05^\circ$ of their positions. This lack of detected X-ray emission is consistent with the negligible blackbody flux estimated from their $E_k$ range and $kT$ adopted in this work.

\end{appendix}
\end{document}